\documentclass[twocolumn,superscriptaddress,showkeys,showpacs]{revtex4}

\usepackage[pdftex]{graphicx}
\usepackage{bm}

\begin{document}

\title[The MAK approach to reconstruction in cosmology]%
{The Monge--Amp{\`e}re--Kantorovich approach to reconstruction in
  cosmology}

\author{Roya Mohayaee}
\email{mohayaee@iap.fr}
\affiliation{Institut d'Astrophysique de Paris, CNRS UMR 7095, 75014
  Paris, France}

\author{Andre{\u\i} Sobolevski{\u\i}}
\email{sobolevski@phys.msu.ru}
\affiliation{Physics Department, M. V. Lomonossov Moscow State
  University, 119992 Moscow, Russia}
\affiliation{International Institute for Earthquake Prediction Theory
  and Mathematical Geophysics, 117997 Moscow, Russia}
\affiliation{Laboratoire J.-V. Poncelet, CNRS UMI 2615, 119002 Moscow,
Russia}

\begin{abstract}
  Motion of a continuous fluid can be decomposed into an
  ``incompressible'' rearrangement, which preserves the volume of each
  infinitesimal fluid element, and a gradient map that transfers fluid
  elements in a way unaffected by any pressure or elasticity (the
  \emph{polar decomposition} of Y.~Brenier).  The Euler equation
  describes a system whose kinematics is dominated by incompressible
  rearrangement.  The opposite limit, in which the incompressible
  component is negligible, corresponds to the Zel'dovich
  approximation, a model of motion of self-gravitating fluid in
  cosmology.

  We present a method of approximate reconstruction of the large-scale
  proper motions of matter in the Universe from the present-day mass
  density field.  The method is based on recovering the corresponding
  gradient transfer map.  We discuss its algorithmics, tests of the
  method against mock cosmological catalogues, and its application to
  observational data, which result in tight constraints on the mean
  mass density \(\Omega_m\) and age of the Universe.
\end{abstract}

\pacs{95.35.+d, 95.75.Pq, 98.62.Py, 98.65.Dx}

\keywords{Dark matter, proper motions of galaxies, reconstruction,
  convex optimization}

\maketitle


\section{\label{intro}Introduction}

In the spectrum of possible models of fluid motion, the Euler equation
of incompressible fluid constitutes an extreme.  As was shown by
Y.~Brenier \cite{Brenier:1987,Brenier:1991}, any Lagrangian motion of
fluid admits a \emph{polar factorization} into a composition of an
``incompressible'' rearrangement, which preserves the volume of each
infinitesimal fluid element, and an ``absolutely compressible''
transfer, which displaces fluid elements to their final locations
prescribed by the gradient of a suitable convex potential, while
expanding or contracting them in a way unaffected by any pressure or
elasticity.  Decomposing a fuid motion into a sequence of small time
steps and factoring out the compressible transfer from the
inertial fluid motion at each step yields a difference scheme for the
incompressible Euler equation \cite{Brenier:1989}.

In this article we show how the opposite approach, in which only the
compressible transfer is retained, can be applied to solving the
problem of reconstructing peculiar motions and velocities of dark
matter elements in cosmology.  We also discuss algorithmics of this
method, which gives an explicit discrete approximation to polar
decomposition and can also be applied to model incompressible fluid as
suggested in \cite{Brenier:1989}.

Recall that on scales from several to a few dozen of \(h^{-1}\)~Mpc
the large-scale structure of the Universe is primarily determined by
distribution of the dark matter.  This distribution can be described
by the mass density field and by the large-scale component of the
peculiar velocity field \footnote{The cosmologial flow of dark matter
  is conveniently decomposed into the uniform Hubble expansion and the
  residual, or \emph{peculiar}, motion.}, controlled by dark matter
itself via gravitational interaction. The dark matter distribution is
traced by galaxies, whose positions and luminosities are presently
summarized in extensive surveys
\cite{Tully:1988,Peacock:2001,Tegmark:2004}.  On large scales
luminosities of galaxies allow to determine their masses, from which
the mass density of the dark matter environment can be estimated using
well-established techniques \cite{Kaiser:1984,Bahcall:2000}.

It is appropriate to consider the reconstruction of the field of
peculiar velocities as part of the more complex problem of
reconstructing the full dynamical history of a particular patch of the
Universe. Several approximate methods have been proposed to this end,
of which we mention here two.  The \textit{Numerical Action Method},
based on looking for minimum or saddle-point solutions for a
variational principle involving motion of discrete galaxies, was
introduced by P.~J.~E.~Peebles in the late 1980s \cite{Peebles:1989};
its modern state is addressed in the present volume by A.~Nusser
\cite{Nusser:2007}.  In this paper we concentrate
on the \textit{Monge--Amp{\`e}re--Kantorovich} method introduced in
\cite{Frisch:2002b} (hereafter the \emph{MAK method}), specifically
highlighting the structural relationship between the MAK method and
the variational approach to the Euler equation of incompressible fluid
\cite{Brenier:2007}.

In Section~\ref{MAK} the mathematical setting of the MAK
reconstruction is derived by application of the Zel'dovich
approximation to a suitable variational formulation of dark matter
dynamics, which leads to the Monge--Kantorovich mass transfer problem
and the Monge--Amp{\`e}re equation.  In Section~\ref{implementation}
we discuss algorithmics of solving the disretized Monge--Kantorovich
problem, which gives as a byproduct an algorithm of polar
decomposition for maps between discrete finite point sets.  In
Section~\ref{testing} we show that the MAK method performs very well
when tested against direct numerical simulations of the cosmological
evolution and review the recent applications of the MAK method to real
observational data, which yielded new tight constraints on the value
of the mean mass density of the Universe.  A detailed treatment of
implementation and testing the MAK method against \(N\)-body
simulations is presented in the companion paper by G.~Lavaux in the
present volume \cite{Lavaux:2007}.  The paper is finished with a
discussion and conclusions.

\section{\label{MAK}Dynamics of cold dark matter and the Zel'dovich
  approximation}

The most widely accepted explanation of the large-scale structure seen
in galaxy surveys is that it results from small primordial
fluctuations that grew under gravitational self-interaction of
collisionless cold dark matter particles in an expanding universe
(see, e.g., \cite{Bernardeau:2002} and references therein).  The
relevant equations of motion are the Euler--Poisson equations written
here for a flat, matter-dominated Einstein--de Sitter universe (for a
more general case see, e.g., \cite{Catelan:1995}):
\begin{eqnarray}
  \label{euler} \partial_\tau\bm v + (\bm v\cdot\nabla_{\bm x}\bm v)
  &=& -\frac 3{2\tau}(\bm v + \nabla_{\bm x}\phi), \\
  \label{continuity} \partial_\tau\rho + \nabla_{\bm x}\cdot(\rho\bm v)
  &=& 0,\\
  \label{poisson} \nabla_{\bm x}^2\phi &=& \frac 1\tau(\rho - 1). 
\end{eqnarray}
Here \(\bm v(\bm x, \tau)\) denotes the velocity, \(\rho(\bm x,
\tau)\) denotes the density (normalized so that the background density
is unity) and \(\phi(\bm x, \tau)\) is a gravitational potential.  All
quantities are expressed in comoving spatial coordinates \(\bm x\) and
linear growth factor \(\tau\), which is used as the time variable; in
particular, \(\bm v\) is the Lagrangian \(\tau\)-time derivative of
the comoving coordinate of a fluid element.  A non-technical
explanation of the meaning of these
variables 
and a derivation of eqs. (\ref{euler}--\ref{poisson}) in the Newtonian
approximation can be found, e.g., in \cite{Brenier:2003d}; see also
\cite{Nusser:2007} in the present volume, where the growth factor is
denoted by \(a\).

The right-hand sides of the Euler and Poisson equations (\ref{euler})
and (\ref{poisson}) contain denominators proportional to
\(\tau\). Hence, it suffices for the problem not to be singular as
\(\tau\to 0\) that
\begin{equation}
  \label{slaving}
  \bm v(\bm x, 0) + \nabla_{\bm x}\phi(\bm x, 0) = 0,\qquad
  \rho(\bm x, 0) = 1.
\end{equation}
Note that the density contrast \(\rho - 1\) vanishes initially, but
the gravitational potential and the velocity, as defined here, stay
finite thanks to our choice of the linear growth factor as the time
variable. Eq.\ (\ref{slaving}) provides initial conditions at \(\tau =
0\); at the present time \(\tau = \tau_0\) the density is prescribed
by a galaxy survey as explained above:
\begin{equation}
  \label{terminal}
  \rho(\bm x, \tau_0) = \rho_0(\bm x).
\end{equation}

In parallel with the Euler equation of incompressible fluid, eq.\
(\ref{euler}) can be considered as the Euler--Lagrange equation for a
suitable action \cite{Giavalisco:1993,Brenier:2003d}:
\begin{equation}
  \label{actionalpha}
  \mathcal I_\alpha = \frac 12\int_0^{\tau_0} d\tau \int d\bm x\cdot
  \tau^\alpha (\rho|\bm v|^2 + \alpha|\nabla_{\bm x}\phi|^2),
\end{equation}
where \(\alpha = \frac 32\) for the flat Universe and minimization is
performed under the constraints expressed by eqs.\
(\ref{continuity}--\ref{terminal}).  Note that the term containing
\(|\nabla_{\bm x}\phi|^2\) may be seen as a penalization for the
nonuniformity of the mass distribution, which corresponds to the lack
of incompressibility of the fluid; enhancing this penalization
infinitely would suppress the ``absolutely compressible'' transfer
of fluid elements, thus recovering the incompressible Euler equation.

However, according to eq.\ (\ref{slaving}) the rotational component of
the initial velocity field vanishes, which strongly suppresses the
``incompressible'' mode of the fluid motion at early times.  Based on
this observation, Ya.~B.~Zel'dovich proposed \cite{Zeldovich:1970} an
opposite approximation in which \(\alpha\to 0\).  In this
approximation eq.\ (\ref{euler}) assumes the form
\begin{equation}
  \label{benamou}
  \partial_\tau\bm v + (\bm v\cdot\nabla_{\bm x})\bm v = 0.
\end{equation}

Much as the incompressible Euler system, the study of the Zel'dovich
approximation benefits from the Lagrangian approach.  Let \(\bm x(\bm
q, \tau)\) be the comoving coordinate at time~\(\tau\) of a fluid
particle that was initially located at~\(\bm q\): \(\bm x(\bm q, 0) =
\bm q\).  Then
\begin{eqnarray}
  \label{jacobian}
  \rho(\bm x(\bm q, \tau), \tau)
  &=& (\det(\partial\bm x/\partial\bm q))^{-1},\\
  \label{velocity}
  \bm v(\bm x(\bm q, \tau), \tau) &=& \partial_\tau \bm x(\bm q, \tau),
\end{eqnarray}
where the \(\tau\) derivative is taken as \(\bm q\) is fixed.  As
observed by Zel'dovich, in these new variables the nonlinear eq.\
(\ref{benamou}) assumes a linear form
\begin{equation}
  \label{lagrange}
  \partial^2_\tau \bm x = 0.
\end{equation}
Moreover eq.\ (\ref{continuity}) is satisfied automatically, and the
action becomes
\begin{equation}
  \label{action0}
  \mathcal I_0 = \frac 12\int_0^{\tau_0}\!\!\!\! d\tau\!\! \int\! d\bm q\,
  |\partial_\tau\bm x(\bm q, \tau)|^2
  = \frac 1{2\tau_0}\int\! d\bm q\, |\bm x_0(\bm q) - \bm q|^2,
\end{equation}
Here we denote \(\bm x_0(\bm q) = \bm x(\bm q, \tau_0)\) and use the fact
that action minimizing trajectories of fluid elements, determined by eq.\
(\ref{lagrange}), are given by
\begin{equation}
  \label{solution}
  \bm x(\bm q, \tau) = \bm q + (\tau/\tau_0)(\bm x_0(\bm q) - \bm q)
\end{equation}
Note that according to the first condition (\ref{slaving}), \(\bm
v(\bm q, 0) = (1/\tau_0)(\bm x_0(\bm q) - \bm q) = \nabla_{\bm q}\bar\Phi(\bm
q)\) and the Lagrangian map (\ref{solution}) remains curl-free for all
\(\tau > 0\): \(\bm x(\bm q, \tau) = \bm q + \tau\nabla_{\bm q}\bar\Phi(\bm q) =
\nabla_{\bm q}\Phi(\bm q, \tau)\) with \(\Phi(\bm q, \tau) = |\bm q|^2/2 +
\tau\bar\Phi(\bm q)\); thus the ``incompressible'' rotational component
of the fluid motion is indeed fully suppressed.

To find the motion of the fluid in the Zel'dovich approximation it is
necessary to minimize the action (\ref{action0}) under the constraint
provided by the representation of density (\ref{jacobian}) and the
boundary conditions (\ref{slaving}, \ref{terminal}):
\begin{equation}
  \label{constraint}
  \det(\partial \bm x_0(\bm q)/\partial\bm q) = 1/\rho(\bm x_0(\bm q)).
\end{equation}
In optimization theory this problem is called the
\emph{Monge--Kantorovich
  problem}
. Equivalently, one can solve the \emph{Monge--Amp{\`e}re
equation} that follows from (\ref{constraint}) for the function
\(\Phi_0(\bm q) = \Phi(\bm q, \tau_0)\) such that \(\bm x_0(\bm q) =
\nabla_{\bm q}\Phi_0(\bm q)\):
\begin{equation}
  \label{mongeampere-q}
  \det(\partial^2\Phi_0(\bm q)/\partial q_i\partial q_j) =
  1/\rho_0(\nabla_{\bm q}\Phi_0(\bm q)).
\end{equation}
At large scales the Lagrangian map \(\bm x_0(\bm q)\) in cosmology is
free from \emph{multistreaming} (the presence of several streams of
dark matter at the same spatial location). Under this assumption the
potential \(\Phi_0(\bm q)\) is necessarily convex \footnote{Convexity
  of the function \(\Phi(\bm q, \tau) = |\bm q|^2 + \tau\Phi_0(\bm
  q)\) holds at \(\tau = 0\) and must be preserved for \(\tau > 0\) if
  no multistreaming occurs.}, and the Legendre transform
\begin{equation}
  \label{legendre}
  \Psi_0(\bm x) = \max_{\bm q}(\bm q\cdot\bm x - \Phi_0(\bm q)),
\end{equation}
where the maximum is attained at \(\bm q\) such that \(\bm x =
\nabla_{\bm q}\Phi_0(\bm q)\), gives (\ref{mongeampere-q}) a simpler
form
\begin{equation}
  \label{mongeampere}
  \det(\partial^2\Psi_0(\bm x)/\partial x_i\partial x_j) = \rho_0(\bm x).
\end{equation}
The \emph{Monge--Amp{\`e}re--Kantorovich (MAK) method}, introduced in
\cite{Frisch:2002b}, consists in solving either of these two problems
for \(\bm x_0(\bm q)\) and using eqs.\ (\ref{solution},
\ref{velocity}) to approximately recover the present field \(\bm v(\bm
x, \tau_0)\) of peculiar velocities.

\section{\label{implementation}Algorithmics of solving
  the~Monge--Kantorovich problem and the~discrete polar decomposition}

To solve the Monge--Kantorovich problem numerically we discretize the
initial and final distributions of mass into collections of Dirac
point masses: all inital point masses \((\mu, \bm q_i)\) are assumed
to lie on a regular grid and be equal, whereas the masses \((m_j, \bm
x_j)\) discretizing the present distribution \(\rho_0(\bm x)\)
typically come from a galaxy survey. The discretized action functional
(\ref{action0}) assumes the form \begin{equation}
  \label{discrete}
  \frac 12\sum_{i, j} \gamma_{ij}|\bm x_j - \bm q_i|^2,
\end{equation}
where \(\gamma_{ij} \ge 0\) denotes the amount of mass transferred
from \(\bm q_i\) to \(\bm x_j\) and mass conservation implies for all
\(i, j\) 
\begin{equation}
  \label{discreteconstraint}
  \sum_k \gamma_{kj} = m_j,\quad \sum_l \gamma_{il} = \mu.
\end{equation}
In practice we choose all \(m_j\) to be integer multiples of the
elementary mass \(\mu\), which guarantees that all \(\gamma_{ij}\)
assume only values \(0\) or~\(\mu\).

Observe that the unknowns \(\gamma_{ij}\) enter into the problem of
minimizing (\ref{discrete}) under constraints
(\ref{discreteconstraint}) linearly.  Problems of this form are called
\emph{linear programs} and can be efficiently solved by various
optimization methods, see, e.g., \cite{Papadimitriou:1982}.  Often the
original, \emph{primal} formulation of a linear program is treated
simultaneously with a \emph{(Lagrange) dual} formulation, which is
another linear program; such algorithms are called \emph{primal-dual}
algorithms.  For the linear program at hand the dual formulation is to
maximize
\begin{equation}
  \label{discretedual}
  -\mu\sum_i \phi_i - \sum_j \psi_j m_j
\end{equation}
under the constraints
\begin{equation}
  \label{discreteconstraintdual}
  \frac 12|\bm x_j - \bm q_i|^2 + \phi_i + \psi_j\ge 0\ \text{for all \(i, j\)}.
\end{equation}
Here \(\phi_i\), \(\psi_j\) are Lagrange multipliers for constraints
(\ref{discreteconstraint}); the duality comes from the following
representation of the (coinciding) optimal values of the two problems:
\begin{eqnarray*}
  \label{duality}
  &&\min_{\gamma_{ij}\ge 0}\max_{\phi_i, \psi_j}
  \biggl(\frac 12\sum_{i,j}\gamma_{ij}|\bm x_j - \bm q_i|^2 \\
  &&\quad + \sum_i\phi_i({\textstyle\sum_j\gamma_{ij}} - \mu)
  + \sum_j\psi_j({\textstyle\sum_i\gamma_{ij}} - m_j)\biggr),
\end{eqnarray*}
where taking \(\max\) or \(\min\) leads to the primal and dual problem
respectively.  The coincidence of the optimal values implies that
\(\bar\gamma_{ij}\), \(\bar\phi_i\), \(\bar\psi_j\) solve the
respective problems if and only if
\begin{equation}
  \label{compslack}
  \sum_{i, j}\bar\gamma_{ij}({\textstyle\frac 12}|\bm x_j - \bm q_i|^2 +
  \bar\phi_i + \bar\psi_j) = 0.
\end{equation}
In view of the nonnegativity conditions this means that for each pair
\((i, j)\) either \(\bar\gamma_{ij} = 0\) or constraint
(\ref{discreteconstraintdual}) is satisfied with equality (and then
\(\bar\gamma_{ij} = \mu\)).  For all other values of \(\gamma_{ij}\ge
0\) and \(\phi_i, \psi_j\) that satisfy constraints
(\ref{discreteconstraint}, \ref{discreteconstraintdual}), the
left-hand side of (\ref{compslack}) is strictly positive and thus the
value of (\ref{discrete}) is strictly greater than that
of~(\ref{discretedual}); such \(\gamma_{ij}\), \(\eta_i, \psi_j\)
cannot be solutions to the respective optimization problems.

A typical primal-dual algorihtm starts with a set of values of
\(\phi_i, \psi_j\) for which all inequalities
(\ref{discreteconstraintdual}) hold and proceeds in a series of steps.
At each step a constraint of the form (\ref{discreteconstraintdual})
is found that is, in a certain sense, the easiest to be turned into
equality, values of the corresponding \(\phi_i\) and \(\psi_j\) are
updated accordingly, and the \(\gamma_{ij}\) is set to~\(\mu\).  An
algorithm stops when the number of equalities in
(\ref{discreteconstraintdual}) equals the number of masses \((\mu, \bm
q_i)\), so that (\ref{compslack}) is satisfied.

We found the \emph{auction algorithm} of D.~Bertsekas
\cite{Bertsekas:1992} (see also \cite{Lavaux:2007}) to be a
particularly efficient primal-dual algorithm for the huge data sets
arising from the cosmological application.  The search for the
constraint (\ref{discreteconstraintdual}) that is to be satisfied with
equality at each step may be performed very efficiently using a
specially developed geometrical search routine, which is based on
suitably modified routines of the ANN library \cite{Mount:2005}.
Further details of this numerical implementation of the MAK method are
given in \cite{Brenier:2003d} and in our forthcoming publication with
M.~H{\'e}non.  A new implementation in a parallel computing environment
is reported in the companion paper of G.~Lavaux \cite{Lavaux:2007}.

We finally show why the solution of the Monge--Kantorovich problem in
the discrete case gives a discrete analogue of polar decomposition.
Let a discrete ``map'' \(\gamma^*\) between two sets of points \((\mu,
\bm q_i)\) and \((m_j,\bm x_j)\) be given such that \(m_j = \sum_i
\gamma^*_{ij}\) and \(\gamma^*_{ij}\) take only values \(0\)
and~\(\mu\).  Solving the corresponding Monge--Kantorovich problem
will give a discrete analogue \(\gamma_{ij}\) of the gradient
transfer.  To see this observe that for \(\Phi_i = \frac 12|\bm q_i|^2
+ \phi_i\), \(\Psi_j = \frac 12|\bm x_j|^2 + \psi_j\) equality in
(\ref{discreteconstraintdual}) means that
\begin{equation}
  \label{eq:1}
  \Psi_j = \max_k (\bm q_k\cdot\bm x_j - \Phi_k)
\end{equation}
(cf.\ (\ref{legendre})), i.e., the map sending \(\bm q_i\) to \(\bm
x_j\) is a discrete analogue of the gradient map \(\bm x(\bm q) =
\nabla_{\bm q}\Phi(\bm q)\) of the previous section.  The
corresponding discrete analogue of ``incompressible'' rearrangement is
a permutation of \((\mu, \bm q_i)\) that for any \(j\) sends the set
of points \(i\) such that \(\gamma^*_{ij} > 0\) to the set of points
\(i'\) such that \(\gamma_{i'j} > 0\); if furthermore \(m_j = \mu\)
for all \(j\), both sets are singletons and the permutation is
recovered uniquely.

In the MAK method we are interested in the ``gradient'' part of the
Lagrangian map sending elements of dark matter to their present
positions, whereas in the difference scheme for the Euler equation
proposed by Y.~Brenier in \cite{Brenier:1989} it is the permutation
part that is retained.

Finally note that an alternative approaches to solve the
Monge--Kantorovich problem, based on direct numerical resolution of
eqs.\ (\ref{benamou}, \ref{continuity}), was proposed in
\cite{Benamou:2000}.

\section{\label{testing}Testing and application of the MAK method to
  cosmological reconstruction}

The validity of the MAK method depends on the quality of the
Zel'dovich approximation, which is hard to establish rigorously.  To
be able to apply the method to real-world data we have instead to rely
on extensive numerical tests.

We report here a test of the MAK method against an \(N\)-body
simulation with over \(2\cdot 10^6\) particles \cite{Mohayaee:2006a}.
The \(N\)-body simulation had the following characteristics: \(128^3\)
particles were assembled in a cubic box of \(200\,h^{-1}\)~Mpc, giving
the mean inter-particle separation of \(1.5\,h^{-1}\)~Mpc; the initial
conditions for the velocity field \(\bm v(\bm q, 0)\) were taken
Gaussian; the density parameter was chosen to be \(\Omega_m=0.3\), the
Hubble parameter to be \(h=0.65\), the normalization of the initial
power-spectrum \(\sigma_8=0.99\), and the mass of a single particle in
this simulation was \(M=3.2\times10^{11} h^{-1} M_\odot\) where
\(M_\odot\) is the solar mass.

\begin{figure}
\includegraphics[width=8cm]{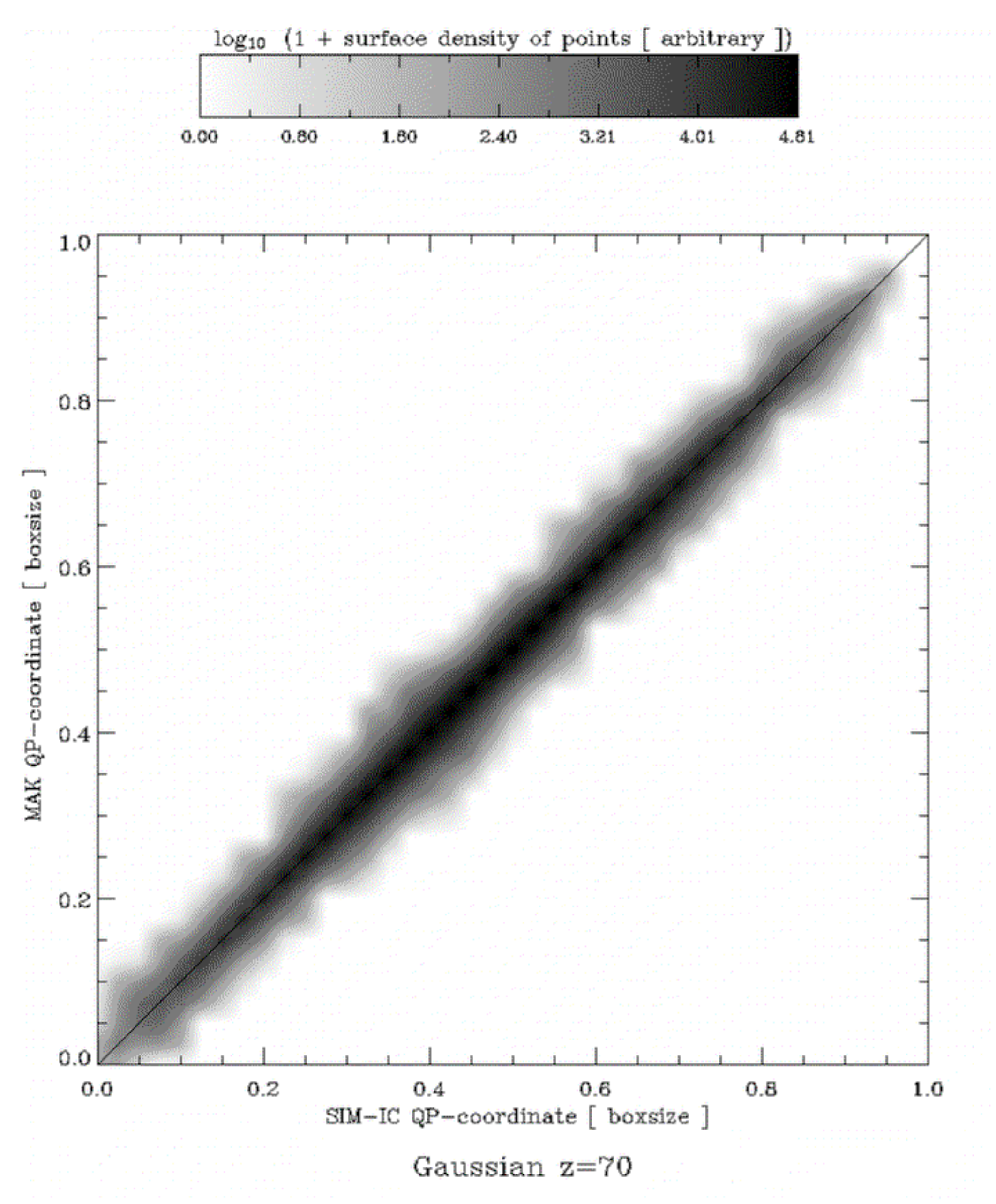}
\caption{Scatter plot of the MAK-reconstructed initial coordinates of
  particles versus their true initial coordinates for a sample of an
  \(N\)-body simulation with \(128^3\) particles in a cubic box of
  size \(200\,h^{-1}\)~Mpc; ideal reconstruction would correspond to
  the diagonal.  A ``quasi-periodic (QP) projection'' coordinate
  \(\tilde q=(q_1+q_2\sqrt 2+q_3\sqrt 3)/(1+\sqrt 2+\sqrt 3)\) is used
  with \(0\le q_i\le1\), where \(1\) corresponds to the (rescaled) box
  size.  The QP projection maps a regular grid in the unit cube into
  the unit segment such that \(\tilde q\) images of no two grid points
  coincide.  Shown is the decimal logarithm of the local density of
  points plus~\(1\); the resolution in QP coordinates is \(1/256\),
  the Lagrangian mesh spacing is \(1/128\).}
\label{fig:scatter-histogram}
\end{figure}

\begin{table}
  \caption{\label{percentages}MAK reconstruction: percentage of
    successfully reconstructed initial positions at different scales
    (measured in units of mesh size in a box of \(128^3\) grid points).}
  \begin{ruledtabular}
    \begin{tabular}{cccccccc}
      Scale & 0\footnote{Exact reconstruction.}
      & \(\le 1\) & \(\le 2\) & \(\le 3\) & \(\le 4\) & \(\le 5\) \\
      \% & 18\% & 41\% & 54\% & 66\% & 74\% & 81\%  \\
    \end{tabular}
  \end{ruledtabular}
\end{table}

\begin{figure}
\includegraphics[width=8cm]{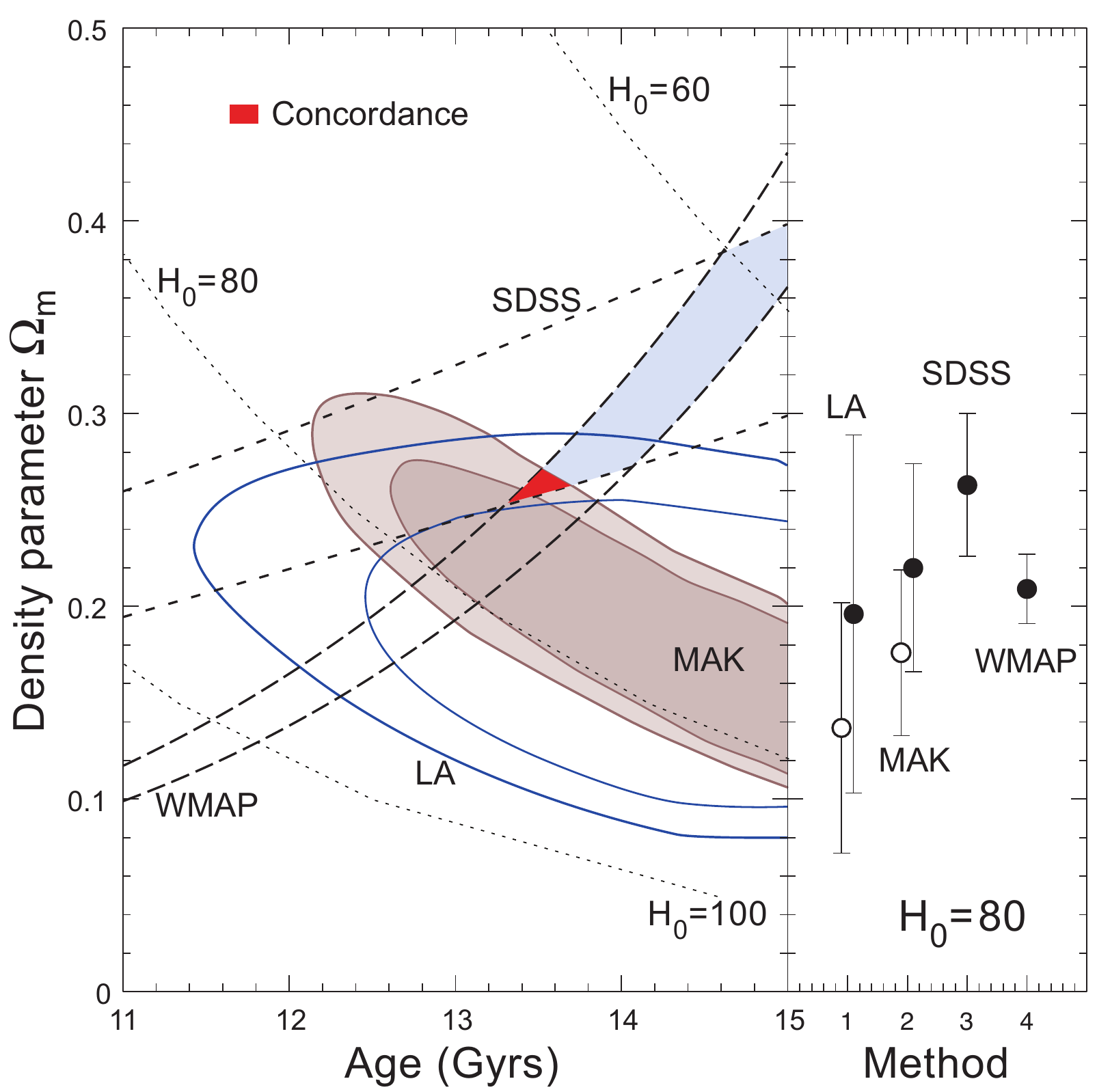}
\caption{\label{fig:parameters} Constraints on the mean mass density
  and the age of the Universe obtained by applying reconstruction
  techniques to real observational data.  Left pane: solid contours
  mark \(\sigma\) and \(2\sigma\) confidence levels for the MAK
  reconstruction (shaded) and the least-action (LA) reconstruction
  (unfilled).  Shaded also is the confluence of the constraints
  on density and age parameters from {\small \emph{WMAP}}
  \cite{Spergel:2003} of \(\Omega_m h^2=0.134\) and from {\small SDSS}
  \cite{Tegmark:2004} of \(\Omega_m h=0.21\).  The  \(2 \sigma\)
  concordance region of the four methods is filled.  
  Right pane: density parameter estimates for \(H_0 = 80\).  For
  details see \cite{Mohayaee:2005a}.}
\end{figure}


The scatter plot in Fig.~\ref{fig:scatter-histogram} demonstrates the
performance of the MAK method in finding Lagrangian positions of the
particles.  At the scales that were probed, positions of about 20\% of
particles are reconstructed exactly (for detailed data see
Table~\ref{percentages}).  This low rate is due to large
multi-streaming at such small scales; at larger scales where mean
inter-particle separation is about \(6\,h^{-1}\)~Mpc (up to \(3\)
meshes), the MAK reconstruction gives the Lagrangian positions of
two thirds of the particles exactly.

At the moment of its implementation in 2006 this reconstruction of
\(128^3\) particles was unprecedented and broke a computational
barrier for cosmological reconstruction schemes; more detailed tests
of the MAK reconstruction of the same scale are reported by G.~Lavaux
\cite{Lavaux:2007} in the present volume.

We now turn to applications of the MAK reconstruction to real
observational data.  The MAK reconstruction of the peculiar velocities
depends on the mean density of the Universe (the parameter
\(\Omega_m\)), the age of the Universe \(\tau_0\) (\(\tau_0\sim
1/H_0\) where \(H_0=100\times h\) is the Hubble parameter), and the
mass-luminosity relation.  Assuming certain values of these
parameters, one can estimate the peculiar velocities of the galaxies
as ratios of their displacements to~\(\tau_0\).  Optimizing the
matching between these velocities and the observed velocities of a few
test galaxies, one can then constrain the parameters of the
reconstruction.

This procedure is illustrated in Fig.~\ref{fig:parameters}, taken from
\cite{Mohayaee:2005a}.  The catalog of galaxies \((m_j, \bm x_j)\)
that is used here is a \(40\%\) augmentation of the Nearby Galaxies
Catalog, now including \(3300\) galaxies within
\(3000\,\mathrm{km}\,\mathrm{s}^{-1}\) \cite{Tully:1988}.  This depth
is more than twice the distance of the dominant component, the Virgo
cluster, and the completion to this depth in the current catalog
compares favourably with other all-sky surveys.  The second
observational component is an extended catalog of galaxy distances (or
radial component of peculiar velocities).  In this catalog, there are
over 1400 galaxies with distance measures within the
\(3000\,\mathrm{km}\,\mathrm s^{-1}\) volume; over 400 of these are
derived by high quality observational techniques that give accurate
estimates of the radial components of peculiar velocities.  The
important feature of Fig.~\ref{fig:parameters} is that the MAK
contours are transversal to contours provided by other methods, which
largely reduces the degeneracy of constraints in the parameter space.


\section{\label{conclusion}Conclusion}

According to the polar decomposition theorem of Y.~Brenier
\cite{Brenier:1987,Brenier:1991}, kinematics of continuous fluid
motion can be decomposed into ``incompressible'' rearrangement and
``infinitely compressible'' gradient transfer.  The Euler equation of
describes a system whose kinematics is dominated by incompressible
motion.  In this paper we show that the opposite limit, in which the
incompressible component is negligible, corresponds to the Zel'dovich
approximation, a physically meaningful model of motion of
self-gravitating fluid arising in cosmology.

This result enables us to approximately reconstruct peculiar
motion of matter elements in the Universe from information on their
present-day distribution, without any knowledge of the velocity
field; indeed, the latter itself can be recovered from the
reconstructed Lagrangian map.  The viability of this method is
established by testing it against a large-scale direct \(N\)-body
simulation of cosmological evolution; when applied to real
obsrvational data, the method allows to get very tight constraints on
the valuse of the mean mass density and gthe age of the Universe.

Another our contribution is an efficient numerical method decomposing
a given displacement field into ``incompressible'' and ``infinitely
compressible'' parts.  This method is not limited to cosmological
reconstruction but can also be used for modelling the dynamics of
incompressible fluid as suggested by Y.~Brenier in \cite{Brenier:1989}.

\begin{acknowledgments}
  This work is partially supported by the ANR grant BLAN07--2\_183172
  (project OTARIE).  AS acknowledges the support of the French
  Ministry of Education and the joint RFBR/CNRS grant 05--01--02807.
\end{acknowledgments}


\end{document}